\documentclass[letterpaper]{article}

\usepackage[T1]{fontenc}

\usepackage{geometry}
\usepackage{setspace}

\usepackage{achemso}

\usepackage{graphicx}
\usepackage{float}
\newfloat{scheme}{htbp}{los}
\floatname{scheme}{Scheme}
\floatname{chart}{Chart}
\newfloat{graph}{htbp}{loh}

\usepackage{chemformula} 
\usepackage[version = 4]{mhchem} 

\usepackage{authblk}
\author[1]{Robin Jacobs-Gedrim*}
\author[1]{William Wahby}
\author[1]{Thomas Awe}
\author[1]{Patrick Xiao}
\author[1]{Melvin Witten}
\author[1]{Jacob Martinez-Marez}%
\author[1]{Kiran Seetala}%
\author[1]{David Hughart}%
\author[1]{Alec Talin}%
\author[1]{Christopher Bennett}%
\affil[1]{Sandia National Laboratories,\\ 1515 Eubank Blvd. Albuquerque, NM, 87123}
\author[2]{Matthew Marinella}
\affil[2]{ Arizona State University, 1151 S. Forest Ave, Tempe, AZ}
\author[3]{Gennadi Bersuker}
\affil[3]{M2D Solutions4620 Trail West Dr., Austin TX, 78735, USA}
\author[1]{Sapan Agarwal}%

\title{Strong radial electric field scaling near nanoscale conductive filaments and the ReRAM resistive switching mechanism}
\date{*Email: rbjaco@sandia.gov}

\begin{document}

\maketitle

\begin{abstract}
 The physics underlying reset in bipolar resistive memory has been the subject of decades of controversy and has been identified as the primary barrier to resistive memory technology development.\cite{dittmann_nanoionic_2021} This manuscript introduces a nanoscale effect in current scaled carrying conductors, whereby surface charge induced radial electric fields are found to be inversely proportional to the radius of the conductive path. This nanoscale effect is then applied to explain the negative resistance switching (reset) mechanism in filamentary metal oxide resistive switching memory devices (memristors).\cite{dittmann_nanoionic_2021} Previous explanations for the negative resistive switching mechanism state that diffusion constitutes the radial driving mechanism for oxygen ions, and drift under electric fields is restricted to the direction parallel to current flow. This explanation conflicts with retention and microscopy data collected in a subset of devices presented in literature.\cite{li_thermodynamic_2024} We demonstrate that the electric field’s dependency on the radius of a nanoscale conductive path can result in radial fields on the order of \(10^5\) to \(10^6\) V/cm at only -1 V bias, sufficient to govern the negative resistive switching mechanism in filamentary metal oxides. By accounting for this novel nanoscale size effect, long-standing anomalous experimental data about the negative (reset) resistance switching mechanism in bipolar filamentary resistive memory devices is finally reconciled. Wide understanding of surface charges and associated electric field in nanoscale conductive paths could prove important for further scaling of integrated circuits and aid in elucidating many nanoscale phenomena.
\end{abstract}

\section*{Keywords}

Resistive Switching, ReRAM, Memristor, Reset, Mechanism, Filamentary, Nanoscale Physics




\section{\label{sec:level1}The current understanding of resistance switching phenomena in filamentary resistive memory}

\begin{figure}[h!] 
    \centering 
    \includegraphics[width=0.9\linewidth]{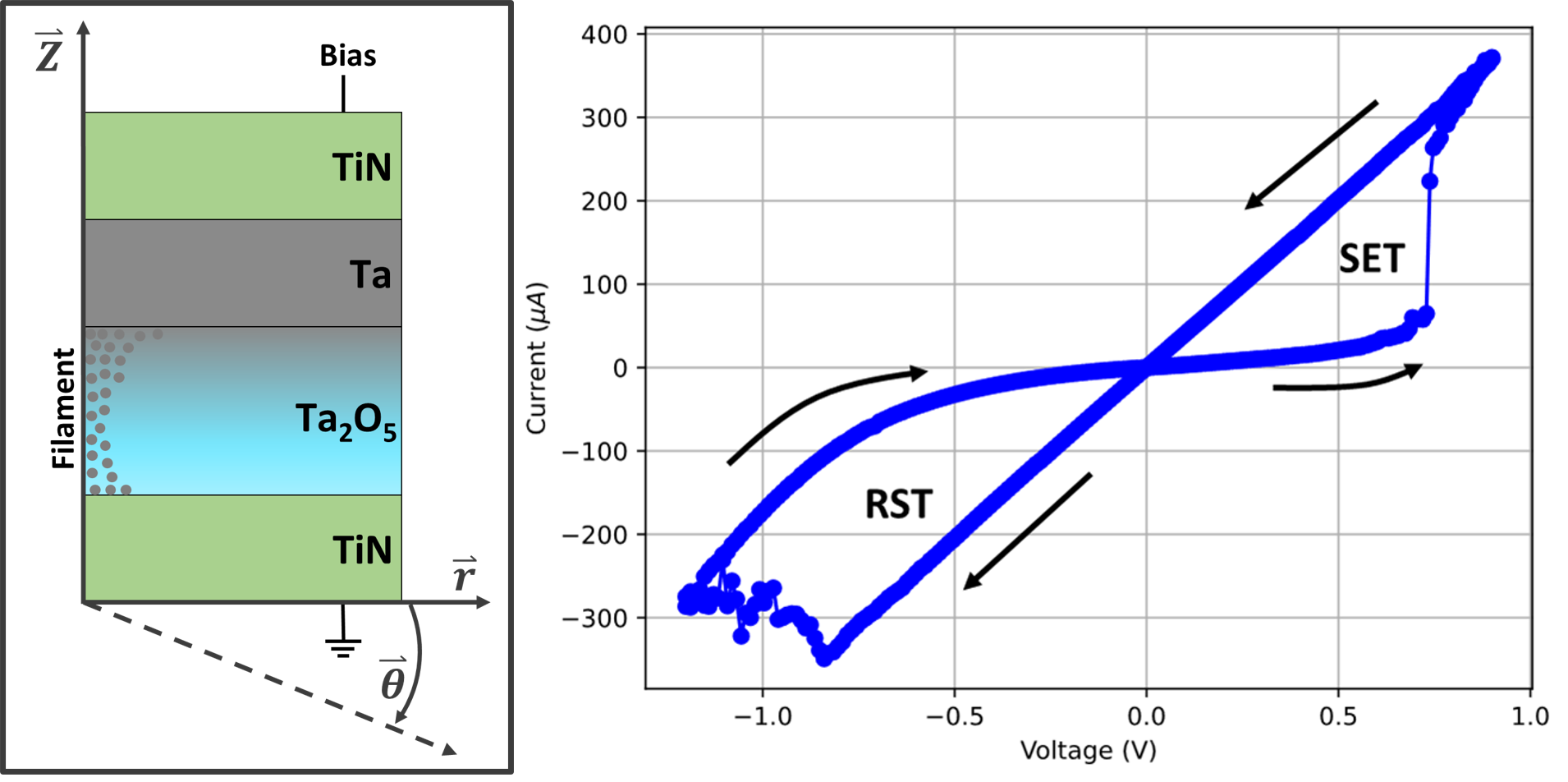} 
    \caption{\(TaO_x\) bipolar Valence Change Mechanism (VCM) ReRAM electrical characteristics. [Top] A typical \(TaO_x\) device is composed of an active Ta electrode Top Electrode (TE), a reduced \(TaO_x\) layer, and an inert TiN bottom electrode (BE). The device can be treated as being cylindrically symmetric around the filament. [Bottom] set-reset  characteristics of the ReRAM device; Positive bias applied to the top electrode (set operation) brings the device to the Low Resistance State (LRS), negative bias applied to the top electrode (reset operation) brings the device to the high resistance state. The set operation is well understood, but the reset operation is still the subject of much controversy.}
    \label{fig:ReRAM_IV_Single} 
\end{figure}

Resistive Random Access Memory (ReRAM) has long been considered a leading candidate for analog weight representation in neuromorphic computing accelerators, compute in memory (CIM), and radiation-tolerant digital non-volatile memory (NVM). Figure \ref{fig:ReRAM_IV_Single} shows the typical electrical current voltage behavior for a \(TaO_x\) ReRAM device during the set-reset switching process. Both forming, the initial soft dielectric breakdown process and the set switch, and the transition to the low resistance state (LRS) in filamentary ReRAM, are  well-described by soft dielectric breakdown mechanisms. Reset, in which the device returns to a high resistance state (HRS) by means of a negative resistance switch, however, remains controversial, motivating a first-principles re-examination of the switching mechanism. \cite{dittmann_nanoionic_2021}\cite{lee_resistive_2015} 

The unresolved reset switching mechanism is said to be the primary obstacle to ReRAM development.\cite{dittmann_nanoionic_2021}\cite{lee_resistive_2015} Early explanations for the reset mechanism in Valence Change Mechanism (VCM) ReRAM assumed an area switching effect based on oxygen ion drift under electric field in the direction parallel to current which was consistent with the bipolar operation principle.\cite{strukov_missing_2008}\cite{dittmann_nanoionic_2021}\cite{lee_resistive_2015} Experimental evidence of the nanoscale filamentary structure of the switching region soon followed, both from microscopy studies and from the lack of resistance scaling with cell area.\cite{sawa_resistive_2008} Additionally, experimental evidence for radial mass transport was shown in Synchrotron Scanning Transmission X-ray Microscopy (STXM) and Conductive Atomic Force Microscopy (C-AFM) experiments. \cite{strukov_thermophoresisdiffusion_2012}\cite{kumar_conduction_2016}\cite{tsurumaki-fukuchi_direct_2023} These data overturned the then-prevailing assumption that electrostatic forces in the direction parallel to current flow—which cannot produce radial mass transport explained the reset mechanism.

Diffusive forces have, thus far, been the only explanation for radial mass transport in VCM ReRAM as shown in Figure \ref{fig:Mechanisms_Single} below. While diffusive forces are clearly present in VCM ReRAM, and many quality models of diffusive forces have been presented, no dynamic simulation model based solely on diffusion has been presented which can consistently model the bipolar reset switch.\cite{kumar_oxygen_2017}\cite{ma_exchange_2020}\cite{mickel_physical_2013} 
\begin{figure}[h!] 
    \centering 
    \includegraphics[width=0.9\linewidth]{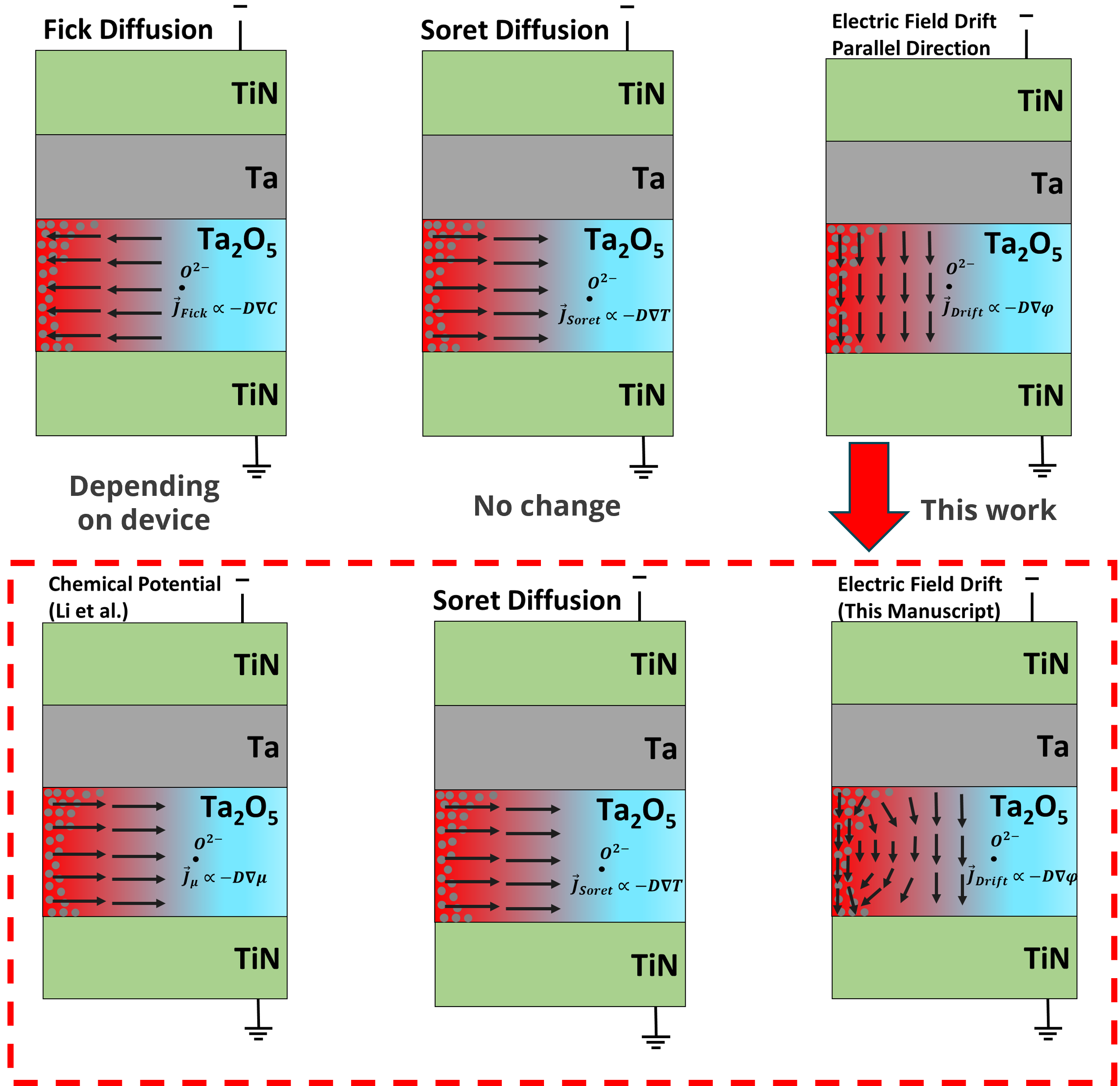} 
    \caption{[Top] Three diagrams of a \(TaO_x\) ReRAM cell during reset with the direction of oxygen ion drift/diffusion indicated according to currently established theory. The center of the filament is located on the left-hand side of each image and the figure is radially symmetric around the filament. SLAC X-ray experiments have shown a ring of excess oxygen which indicated a radial reset switching mechanism invalidating earlier electric field-only switching mechanisms. Currently, the only drift/diffusion vector with the correct direction to explain the reset mechanism is thought to be Fick diffusion. (Bottom) Three diagrams of a \(TaO_x\) ReRAM cell during reset with the direction of oxygen ion drift/diffusion indicated according the theory presented in this manuscript. Recent experiments have shown that in \(TaO_x-TaO_y\) the direction of the chemical potential is reversed from the concentration gradient, so there is no diffusive mechanism that can explain reset in this subset of devices. TEM experiments have shown that the filament consistently breaks near the Bottom Electrode (BE), but there is not a simple explanation for this behavior in established theory.}
    \label{fig:Mechanisms_Single} 
\end{figure}

Thermodiffusion (Soret effect), which arises from the net effect of having high stochastic molecular movement in a hot region and low molecular movement in a cool region, leads to an overall molecular flow from the hot region to the cool region. During the reset process in VCM ReRAM, Joule heating in the filament produces a temperature gradient. Free oxygen ions\textbf{*} are then driven from the hot interior of the filament to the cool surrounding region. However, thermodiffusion cannot fully account for reset, since it should drive oxygen out of the filamentary region, lowering resistance. 

Diffusion of oxygen down the oxygen concentration gradient (Fick diffusion) does predict an increase in resistance, however a recent study has shown that the amorphous tantalum oxide system exhibits uphill diffusion due to the miscibility gap between metallic tantalum rich phases and insulating oxygen rich phases.\cite{li_thermodynamic_2024} Diffusion-only-models conflict with recent data of retention and failure mechanisms in bipolar ReRAM as they sometimes predict a temperature limit to the LRS that is exceeded in experimental data, and a collapse to the HRS which is not observed in all devices.\cite{li_thermodynamic_2024} 

STEM imaging of the filament in bipolar conductive filament ReRAM reveals that rupture consistently occurs near the bottom electrode (BE) even when an hourglass filament is observed, which contradicts expectations for a diffusive mechanism. \cite{arita_switching_2015}.  Rupture near the BE is well accounted for in conical filaments, and while a conical filament is a common shape, other morphologies are often visible in STEM experiments such as hourglass filaments or the morphology is not clear.\cite{wang_conduction_2018} \cite{ma_stable_2019} Finite element based multiphysics simulations taking into account all physical forces have successfully modeled the negative resistance (reset) switch in filamentary ReRAM\cite{kim_comprehensive_2014}\cite{la_torre_compact_2019}\cite{niraula_numerical_2017} but there still exists an understanding gap between the computer simulation and human interpretation/confidence in results due to the large number of parameters involved. In Multiphysics models, field dependent drift is shown to dominate the switching process  but the lack of a drift-based explanation for reversible radial mass transport has prevented drift from being accepted as the primary switching mechanism.\cite{a_marchewka_physical_2017}\cite{lee_quantitative_2020} 

Radial fields are most surprising for a metallic or degenerately doped semiconducting filament, and the burden of proof the highest for this case, so that was to be the focus of this paper. As reset is reproducible under DC and quasi-static bias, it can be treated as a steady-state electrostatic phenomenon. While electrodynamic effects and skin effect will come into play during the resistance change in the RESET switch, or when using high frequency pulses to create a RESET switch, they are omitted here for simplicity. This manuscript will demonstrate that a re-examination of radial electric fields around the filament can reconciles the discrepancies between theory and experiment regarding the reset resistance switching mechanism.
\\

\textbf{*}Please note: This paper will discuss the movement of oxygen ions rather than the commonly used oxygen vacancies term, as it simplifies the particular discussion, and the use of the term oxygen vacancies is contentious in amorphous materials. In all cases analyzed, oxygen vacancies travel in the opposite direction of oxygen anions.

\section{\label{sec:level1}Surface charges and radial electric fields}
The electric field inside and outside of current carrying conductors with finite resistance is generated by a buildup charge at the surface of the conductive path was originally described by Kirchhoff in 1857.\cite{graneau_kirchhoff_1994} An excellent historical discussion on why this fact is not more widely recognized today can be found in .\cite{assis_electric_1999} While radial electric fields near macroscale wires are weak, convincing experimental evidence for their existence can be found in high current experiments by Jefimenko and in exploding wire pulsed power experiments by Sarkisov et. al.\cite{jefimenko_electricity_1966}\cite{sarkisov_inverse_2018}. The following is proof that there are radial electric fields outside of current carrying conductors, and that there is charge located at the surface of the conductive path.\cite{russell_surface_1968} The Drude model shows that there is an electric field inside a steady state current carrying conductor. From the classical version of the Drude model, the current density \( J \) is given by

\begin{equation}
J = \frac{nq^2 \tau}{m} E
\end{equation}

where \( J \) is the current density, \( m \) is the electron mass, \( n \) is the number of charge carriers, and \( \tau \) is the mean free time between ionic collisions. As current flows in the conductor, there is an electric field in the interior of the conductor. As there are no sources of emf and no volume distributions of charge, the potential functions in the conductor and near the conductor’s surface must satisfy Laplace’s equations. Assuming steady currents and uniform resistance between terminals, the divergence of current is zero in the interior of the conductor, such that

\begin{equation}
\nabla \cdot J = \frac{nq^2 \tau}{m}\quad\nabla\cdot E = 0
\end{equation}

Thus, the electric field is constant everywhere in the interior of the conductor.\cite{jackson_classical_1999}\cite{griffiths_introduction_2023} The direction of current flow, aligned with the \( z \) direction, will be parallel with current flow. Assuming steady currents and uniform fields, it is possible to make the assertion that \(E_z = B\), where \( B \) is a constant in the interior of the conductor. Due to symmetry, the tangential electric field is zero \(E_\theta = 0\). Due to the Lorentz force 

\begin{equation}
F = q(E + v \times B),
\end{equation}

there is some radial electric field. However, due to the low drift velocity and low magnetic fields given the mA currents and \( B \propto \frac{1}{r} \) scaling of the magnetic field, the radial electric field in the conductor becomes negligible such that \(E_r \cong 0\). A full treatment of the transverse electric fields in current carrying conductors can be found in  \cite{matzek_transverse_1968}. Given a constant electric field \(E_z=B\), we can apply Gauss’s law, \(V_z=-\int{E\cdot d z}\) and integrate the constant to show that the potential in the interior of the conductor is a linear function of z, such that \(V_z=Z\left(z\right)=\left(A+Bz\right)\). This generalizes to 

\begin{equation}
V\left(r,\theta,z\right)=\left(A+Bz\right)\cdot F\left(r,\theta\right)
\label{eq:Russel}
\end{equation}

in the region around the conductor by noting the separability of the z coordinate.  There is no theta dependence to this potential.  At the surface of the conductor, where r=R, the potential is entirely a function of the z coordinate: \(V\left(z\right)=\left(A+Bz\right)\). Taking the derivative of potential just inside of the conductive cylinder's surface, and the derivative just outside of the surface, shows a discontinuity in the derivative of potential, as it is zero inside the cylinder and non-zero outside. Any region over which the derivative of the potential has a discontinuity must contain some charge from 

\begin{equation}
\varepsilon_{outisde}{\mathrm{\nabla V}}_{outside}-\varepsilon_{inside}\mathrm{\nabla}V_{inside}=\sigma\hat{r}
\end{equation}

thus there is charge located at the surface of the cylinder, \(\sigma\left(R,\theta,z\right)=A+Bz\). As noted, \({\mathrm{\nabla V}}_{outside}\neq0\) so there are radial electric fields outside any conductive path with classical conduction mechanisms and finite resistance Q.E.D.

\begin{figure}[h!] 
    \includegraphics[width=0.8\linewidth]{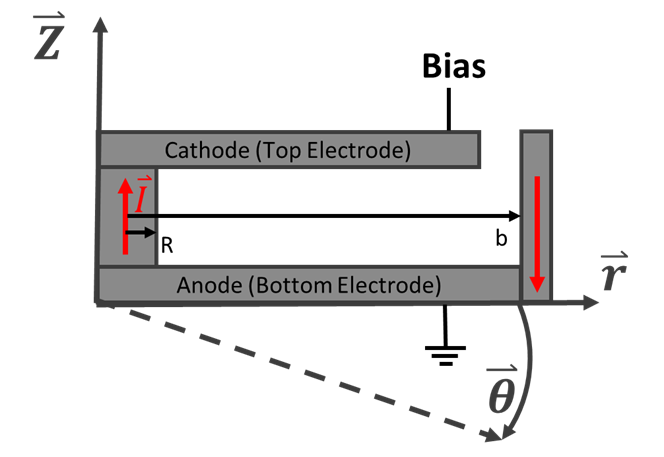} 
    \centering 
    \caption{The boundary conditions for a simplified resistive memory cell with a cylindrical filament in a metal insulator metal (MIM) capacitor structure with negative bias applied to the top electrode and ground on the bottom electrode is shown above. The ground path is assumed to be a V=0, at a distance b from the filament in the r direction, which is \(>>\) the radius of the top and bottom electrodes. The radius of the filament is R and is aligned with the R direction. This system is fully cylindrically symmetric in \(\theta\).}
    \label{fig:Boundary_cond} 
\end{figure}

The form and magnitude of the electric field depends strongly on the boundary conditions and location of the return path. The following derivations follow the boundary conditions and solutions to Laplace’s equation originally intended to examine radial electric fields; this caused microscale wires to explode in pulsed power experiments.\cite{awe_experimental_2009}  A diagram of the filament sitting in a capacitor-like structure, with a cylindrical return path sitting at distance $b$ from the filament is shown in Figure \ref{fig:Boundary_cond}. 

From the Drude model, and collecting constants, current density inside the cylinder is \(E_z=\eta J_z\). Thus, the electric field in the z direction is of the \(E_z=\frac{\eta I}{\pi R^2}\) Where \(\eta\) is resistivity and \(J_z\) is the current density in the z direction. Thus, the electric field in the z direction is of the \(E_z=\frac{\eta I}{\pi R^2}\) Where \(\eta\) is resistivity and \(J_z\) is the current density in the z direction. Voltage at the surface of the filament (for \(r=R\)) is just the integral of the electric field in the z direction \(V\left(R,z\right)=-\int_{0}^{z}{E_z\left(R\right)\cdot d z}= -\frac{\eta Iz}{\pi R^2}\). To solve for the electric field outside of the filament, it is possible to apply Laplace’s equation: 

\begin{equation}
\mathrm{\nabla}^2V=\frac{1}{r}\frac{\delta}{\delta r}\left[\frac{\delta V}{\delta r}\right]+\frac{1}{r^2}\frac{\delta^2}{\delta\theta^2}+\frac{\delta^2V}{\delta Z^2}=0. 
\end{equation}

Noting the lack of \(\theta\) dependence, Laplace’s equation reduces to, \(\mathrm{\nabla}^2V=-\frac{1}{r}\frac{\delta}{\delta r}\left[r\frac{\delta V}{\delta r}\right]+\frac{\delta^2V}{\delta Z^2}=0\). Plugging Equation \ref{eq:Russel} into the Laplacian gives \(\frac{1}{r}\frac{\partial}{\partial r}\left(r\frac{\partial V}{\partial r}\right)=\frac{1}{r}\frac{\partial}{\partial r}\left(r\frac{\partial}{\partial r}\left(zf\right)\right)=\frac{z}{r}\frac{\partial}{\partial r}\left(r\frac{\partial f}{\partial r}\right)=0\), which reduces to the ordinary differential equation 

\begin{equation}
\frac{z}{r}\frac{\partial}{\partial r}\left(r\frac{\partial f}{\partial r}\right)=0
\end{equation}

This can be evaluated using the product rule, \(\frac{\partial}{\partial x}\left(g\left(x\right)\cdot h\left(x\right)\right)=g\frac{\partial h}{\partial x}+\frac{\partial g}{\partial x}h\), so \(\frac{z}{r}\frac{\partial}{\partial r}\left(r\frac{\partial f}{\partial r}\right)=\frac{z}{r}\left(r\frac{\partial^2f}{\partial r^2}+1\cdot\frac{\partial f}{\partial r}\right)=\frac{z}{r}\left(r\frac{\partial^2f}{\partial r^2}+\frac{\partial f}{\partial r}\right)=0\). Collecting terms and again noting separability of z and r gives \(\frac{z}{r}\left(r\frac{\partial^2f}{\partial r^2}+\frac{\partial f}{\partial r}\right)=z\left(\frac{\partial^2f}{\partial r^2}+\frac{1}{r}\frac{\partial f}{\partial r}\right)=0\). The result is the product of two equations with different variables, \(z\left(\frac{\partial^2f}{\partial r^2}+\frac{1}{r}\frac{\partial f}{\partial r}\right)=p\left(z\right)\cdot q\left(r\right)=0\), so the only way for this equation to equal zero is for the r-dependent portion to cancel out, leaving the simple ordinary differential equation (ODE):

\begin{equation}
\frac{\partial^2f}{\partial r^2}=-\frac{1}{r}\frac{\partial f}{\partial r}. 
\end{equation}

The equation \(F\left(r\right)=Aln\left(B\cdot r\right)\) satisfies the ODE which can be proved by substitution. \(\frac{\partial^2\ln(r)}{\partial r^2}=-\frac{1}{r}\frac{\partial\ln{\left(r\right)}}{\partial r}\  =>  \frac{\partial}{\partial r}\left(\frac{1}{r}\right)\ =-\frac{1}{r}\cdot\frac{1}{r}\) so \(-\frac{1}{r^2}=-\frac{1}{r^2}\). Assuming a zero potential return path at r=b larger than the capacitor structure in which the switchable film is located \(V(b,z)=0\) gives \(B=(1/b)\), \(V\left(R,z\right)=z\cdot F\left(R\right)=z\cdot Aln\left(B\cdot R\right)=-\frac{\eta Iz}{\pi R^2}\Rightarrow A= -\frac{\eta I}{\pi R^2}\left[ln\left(\frac{R}{b}\right)\right]^{-1}\). Therefore the potential function is: 

\begin{equation}
V\left(r,z\right)=-\frac{\eta Iz}{\pi R^2}\left[ln\left(\frac{R}{b}\right)\right]^{-1}\left[\ln\left(\frac{r}{b}\right)\right]. 
\end{equation}

Taking the derivative with respect to r or z gives the electric field in these directions:

\begin{equation}
{E\left(r,z\right)}_r=-\frac{\delta V\left(r,z\right)}{\delta r}=\frac{\eta Iz}{\pi R^2} \frac{1}{r\cdot\ln(R/b)}.
\end{equation}

\begin{equation}
{E\left(r,z\right)}_z=-\frac{\delta V\left(r,z\right)}{\delta z}= -\frac{\eta I}{\pi R^2}\frac{\ln(r/b)}{\ln(R/b)}
\end{equation}

In a nanoscale system the number of conductive paths and the individual conductivity of those paths may not be simple to determine, so it is useful to employ Ohm’s law for a parallel resistors \({E_n\left(R,z\right)}_r= \frac{\eta z\sum_{1}^{N}I_n}{\pi R^3} \frac{1}{ln(R/b)}\ =\ \frac{{(zR}_{res}\pi R^2/l)\sum_{1}^{N}I_n}{\pi R^3}\frac{1}{ln(R/b)}=\frac{Vz}{Rl}\frac{1}{ln(R/b)}\). Therefore, the radial electric field near a sub-filament in a dielectric material with dielectric constant k is 

\begin{equation}
{E_n\left(R,z\right)}_r =\frac{Vz}{kRl\cdot\ln(R/b)}
\end{equation} regardless of the total number of conductive paths or the individual conductivity of those paths. The radial electric field in the above equation increases inversely proportional to the radius of the conductive path; if the conductive path is nanoscale in dimension, significant surface charge and strong radial fields may be present at modest bias.

\section{Calculation of Radial Electric Field Strength Near Nanoscale Filaments}

Determination of the filamentary structure of Valence Change Mechanism (VCM) based ReRAM is difficult due to the small size of the filament, and the fact that the filamentary region is composed of the same elements as the non-conductive region, just at different stoichiometry, and the fact that the filament is buried within metal electrodes and can appear in a random location withing a larger electrode area. Top-down area measurements such as conductive atomic force microscopy (C-AFM), Electron Beam Induced Current (EBIC), X-ray spectromicroscopic techniques, pressure modulated conductance microscopy (PMCM) etc. often report filamentary regions of 10-40 nm in radius.  These results may overestimate the size of the filament as they report the largest radius of the filamentary area, require significant over current during forming to make the filament visible, or often occlude complex substructure or the conglomeration of smaller sub filaments.\cite{celano_imaging_2015}Metrology techniques which measure a cross section of the filament, such as Scanning Transmission Electron Microscopy (STEM), or techniques which produce a 3D map of the filament such as Scalpel C-AFM, have shown that underneath the 20-50 nm filamentary region, there are one or more filaments with radius found to be between 0.5 nm and 7.5 nm in \(TaO_x\) and even in this size range, the presence of smaller sub filaments has been confirmed by both scanning probe and transmission electron microscopy.\cite{celano_imaging_2015}\cite{park_situ_2013}\cite{ma_stable_2019} 

\begin{figure}[t] 
    \centering 
    \includegraphics[width=0.9\linewidth]{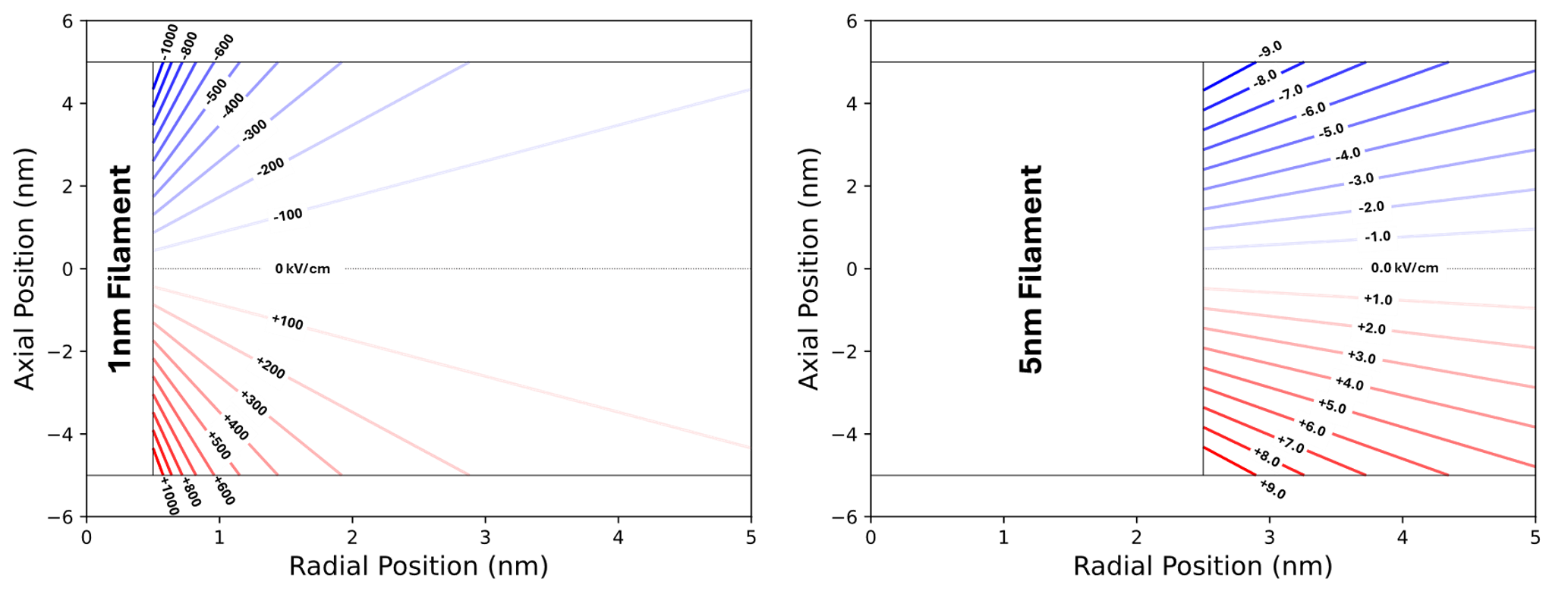} 
    \caption{Analytical solution for the radial component of the electric field near a 1 nm diameter (left) and 5nm diameter (right) conductive path, with a distance to the return path b=500 nm, 1 mA current on the filament, and a filament height of 10 nm. The radial component of the electric field is significantly stronger for thinner filaments.}
    \label{fig:Analytical_E_Fields_Single.png} 
\end{figure}

We use the smaller range of estimates for filament radius, between 0.5 nm and 5 nm. The calculated radial electric fields for a ReRAM filament using the 1nm and 5nm diameter conductive paths (0.5nm and 2.5nm diameters), with a distance to the return path b=500 nm, 1 mA filament current, and a filament height of 10 nm are shown in Figure \ref{fig:Analytical_E_Fields_Single.png}. 
The radial electric field is significantly stronger for thinner filaments, with the 1 nm filament exhibiting field intensities in the \(10^5-10^6\) V/cm range, compared to \(10^3-10^4\) V/cm for the 5 nm filament. Electric fields of \(10^5-10^6\) V/cm magnitude can both mediate electron emission and cause chemical bonds breakage that may result in dielectric breakdown.

\section{Finite element model of the electric field in filamentary resistive memory}

The analytical calculations assume a generic conductor and are limited to cylindrical filaments. Additionally, the derivation in the analytical model requires a grounded return path located at a distance b from the filament, which is not physically present in most ReRAM devices. To demonstrate and further confirm that radial electric fields are present a finite element model for a conductive path was produced using COMSOL Multiphysics software. In Figure \ref{fig:Finite_Element_Single.png} below, electric fields for a 1 nm and 5 nm radius cylinder 10 nm in z dimension are shown. 
\begin{figure}[h!] 
    \centering 
    \includegraphics[width=0.9\linewidth]{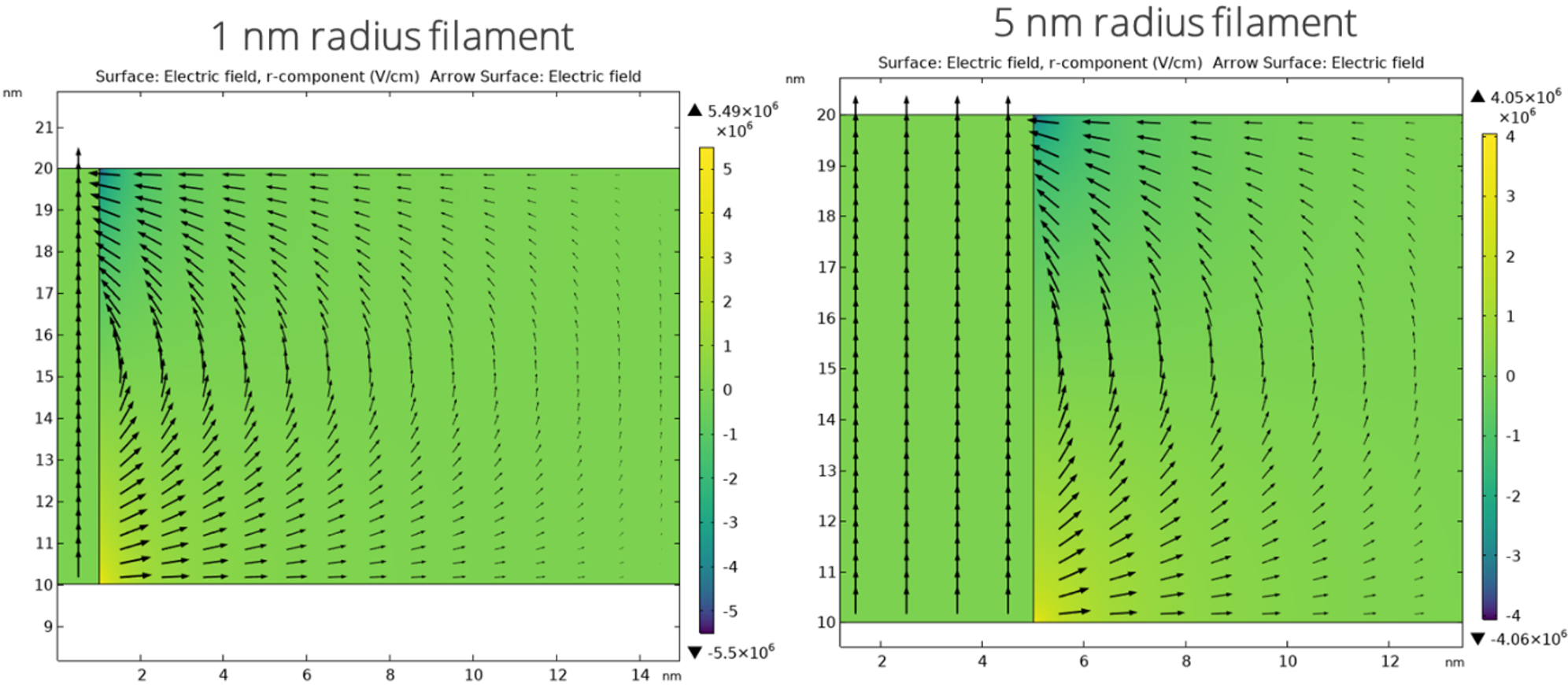} 
    \caption{Finite element model of radial electric field for a cylindrical tantalum filament of 1 nm and 5 nm respectively under application of -1V bias to the top of the cylinder and ground applied to the bottom of the cylinder. Black arrows show the electric field direction and their length is proportional to magnitude. Radial electric fields reach 5.49 MV/cm near the 1 nm filament surface which is stronger than the dielectric strength of \(Ta_2O_5\), and radial fields reach 4.05 MV/cm for the 5 nm filament. These fields may be enhanced above the analytical expressions due to the sharp corners in the model. The vector direction of the radial field is critical, as negatively charged oxygen ions move towards the bottom of the filament where it has been shown to rupture in experiment.}
    \label{fig:Finite_Element_Single.png} 
\end{figure}
The insulating region was modeled as stoichiometric \(Ta_2O_5\) with a conductivity of \(8\cdot10^{-10}\) S/m and a dielectric constant of 25. Meanwhile, the conductive path was modeled as tantalum with a conductivity of \(1.59\cdot10^{2}\) S/m. The model is cylindrically symmetric surrounded by electric insulation. A potential of -1 V was applied to the top surface of the cylindrical filament and 0 V was applied to the bottom electrode. Both an electric field of necessary magnitude to potentially free oxygen, and the necessary radial direction is observed to facilitate the RESET mechanism is observed in the finite element model. The peak positive electric fields which indicate rupture location are always near the bottom electrode. This explains why the filament is observed to rupture near the bottom electrode without requiring that the filament have any particular morphology.

This study establishes that ReRAM exhibits a surface charge dependent resistive switching mechanism in the reset direction. Surface charges are the source of the electric field which drives current through a conductive path. Any conductor carrying a current is a source of magnetic fields from the Biot-Savart law.\cite{jefimenko_electricity_1966}\cite{jackson_classical_1999}\cite{griffiths_introduction_2023} Therefore, there is a connection between surface charge, changing resistance, and magnetic flux linkage.\cite{Chua_Memristor_1971} 

\section{Conclusions}

This paper demonstrates that that strong radial electrostatic forces arise during reset due to surface charges, and along with thermal diffusive forces may complete the theory of resistive switching in metal oxides. The surprising strength of the electric field generated by a current traveling through the nanoscale conductive path is elucidated, and an explanation for the negative resistance switching (reset) mechanism in filamentary resistive memory is demonstrated. Joule heating, thermal diffusion (Fick/Soret), and charge injection are expected to still play a role in freeing oxygen and in the dynamics of switching and the evolution of the filament morphology. This radial electric field effect elucidates for the first time several key features of experimental evidence which have not previously had adequate theoretical explanations: 1.) the first explanation for bipolar switching which works for all devices 2.) how radial mass transport can arise around a filament in a drift dominated system 3.) the rupture of the ReRAM filament being preferentially located near the bottom electrode  4.) retention in high temperature environments and 5.) the reason that the filamentary resistance switching mechanism is so material system independent. Understanding the size effect introduced here will enable engineering of reliable operational characteristics of resistive memory devices and will be important for scaling electronics down to single digit nanometer dimensions.

\section{Acknowledgments}

The authors would like to thank Y. Li, A. Mitchell, F. Doss, C. James, G. Sarkisov, J.A. Incorvia, A. Dozier, J. Hasler,  and E. Tutuc for the useful discussions. We would also like to thank C. Carlos, P. Finnegan, for experimental work that lead to ideas contained in this paper. Special thanks to G. Haase for serving as the internal referee. This paper is dedicated to S. Williams, L. Chua, and colleagues, who inspired our interest in this field.
\\

This paper describes objective technical results and analysis. Any subjective views or opinions that might be expressed in the paper do not necessarily represent the views of the U.S. Department of Energy or the United States Government.
\\

Sandia National Laboratories is a multimission laboratory managed and operated by National Technology and Engineering Solutions of Sandia, LLC, a wholly owned subsidiary of Honeywell International Inc., for the U.S. Department of Energy’s National Nuclear Security Administration under contract DE-NA0003525. SAND\#0000-XXXXX

\bibliography{Radial_Field_Paper_bibtex}

\end{document}